# Measure of precursor electron density profiles of laser launched radiative shocks

Michel Busquet (1), Frédéric Thais (2), Matthias González (3), Edouard Audit (3)

[1] ARTEP, inc, Ellicott City, MD, USA

[2] CEA Saclay/DSM /IRAMIS/SPAM, 91191 Gif-sur-Yvette Cedex, France

[3] Laboratoire AIM, Service AstroPhysique, CEA Saclay/DSM/IRFU-CNRS-Université Paris Diderot, F-91191 Gif-sur-Yvette Cedex, France

**Corresponding author** : Michel Busquet, ARTEP, inc, Ellicott City, MD, USA

**mail** : busquet@artepinc.com

**Abstract**

We have studied the dynamics of strong radiative shocks generated with the high-energy sub-nanosecond iodine laser at PALS facility (Prague) over long time scales, up to 100 ns. These shock waves are characterized by a developed radiative precursor: a radiation driven ionization wave in front of the density jump of the shock. Electronic density profiles are measured at different times after the laser pulse and at different distances from the axis of the shock tube. A new feature, described as a split precursor, has been observed. Comparisons with 2D computations are shown.

# 1 – INTRODUCTION

Radiative shocks [1] are strong shocks where the flux of ionizing photons coming from the shock front is large enough to launch an ionization wave in front of the shock. This radiative precursor heats the medium in front of the density jump to a temperature often equal to the temperature of the shocked medium. Radiative-precursor shocks are relevant to astrophysics, in supernova [2], supernova remnants [3], accretion shocks [4] and jets [5]. A criterion for a velocity threshold of radiative precursors has been derived balancing downstream ionizing photons and upstream atoms being ionized [6]. Radiative shocks have been produced in the laboratory at laser facilities with blast waves in residual gas inside the vacuum chamber [7], or with shock waves in foam [8], or in gas filled mini-shock tubes [9,10,11], where the shock is launched by a laser ablated foil acting as a piston. Scaled down to the laboratory, as the attainable velocity is ruled by the available laser energy, the velocity threshold criterion lead to use high atomic number gas, Argon or Xenon, at low pressure (0.05 – 5 bar). Studies of radiative shocks in Xenon have been initiated on the CEA-Limeil facility [12], followed by experiments on the LULI facility [9] and on the Omega facility [10]. Recently we achieved the first long time observation (over 40 ns) on the PALS facility [11]. In the study we are presenting here, time of analysis is even larger, as we observe the radiative precursor up to 100 ns and we measure electron density profiles up to 50 ns, with various plasma conditions and observation geometries. Preliminary results have been presented elsewhere [13], in this paper we go into more detailed analysis, including synthetic interferogram, and complete or correct some values, leading to a refined conclusion.

# 2 – EXPERIMENTAL SETUP AND SOME THEORY ABOUT RADIATIVE SHOCKS

The experiment was performed at the Prague Asterix Laser System facility (PALS) [14]. The experimental setup (Fig.1) includes two main diagnostics : a time resolved interferometer using a green laser ($\lambda=0.527$ μm) to probe the electron density in the radiative precursor and a time-integrated backside XUV spectroscopy. The frequency tripled ($\lambda_3$ =0.438 μm) laser pulse of the PALS iodine laser is focused on the target, a miniaturized shock tube described below. The nominal pulse conditions are a 0.35 ns duration and a 150-170 J energy (in green) after the infrared rejecting filter. The 29 cm diameter PALS beam is smoothed by a Phase Zone Plate (PZP) [15], located before the main PALS focalization lens. The PZP yields a flat top focal spot of 0.55 mm FWHM in diameter, with 90% of the energy in a 0.7 mm circle, matching the inner section of the smallest shock tubes. Accounting for an incomplete transmission for each of the PZP, the focusing lens and the vacuum window, the resulting energy on target is 110-130 J. Then mean laser intensity on target is $10^{14}$ W/cm$^2$ in green ($\lambda_3$=0.438 µm).

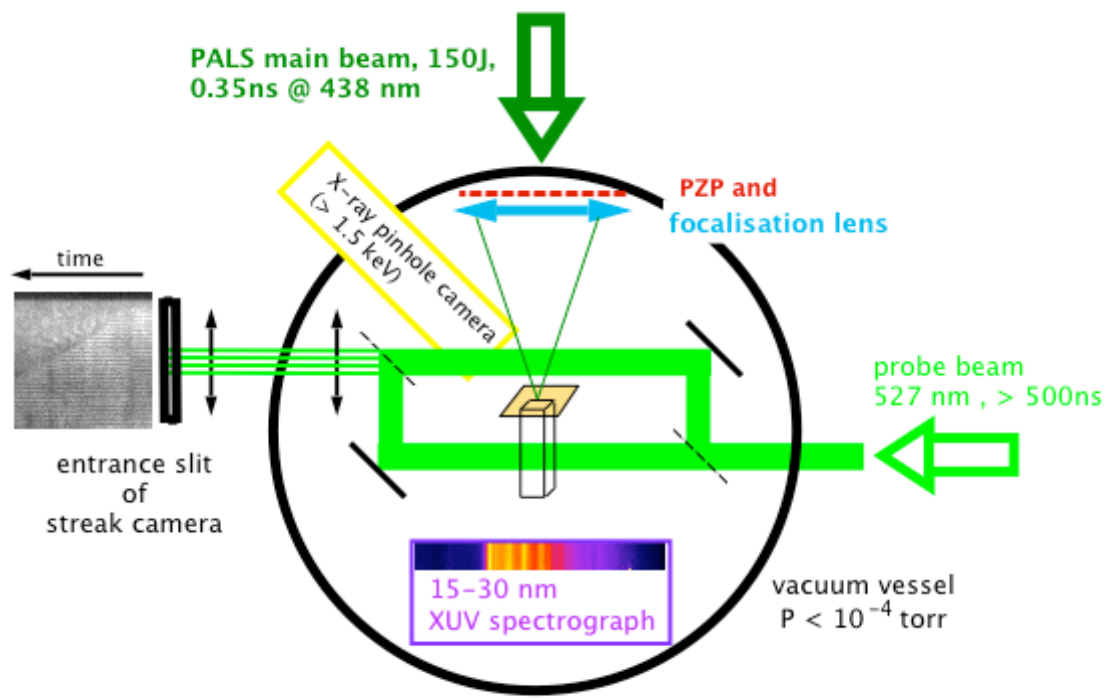

*Figure 1 : experimental setup (see text for details).*

We use miniaturized shock tubes (Fig.2) filled with Xenon at low pressure (or air-Xenon mixtures in some case). The specially designed targets [16] were manufactured at "Pôle Instrumental de l'Observatoire de Paris". Two geometries have has been used :

- (a) tubes of square internal section, with 2 metallic sides and 2 lateral glass windows (all 4 sides can be gold or aluminium coated) allowing longitudinal and transverse interferometric study (see description below)

- and (b) blind metallic cylindrical tubes (without any lateral window) for back side XUV spectroscopy in the direction of the shock propagation..

All the targets are closed on the top by a thin foil. This foil is a composite film, manufactured at IPN Orsay, made from a 10 µm polystyrene foil with a 0.5 µm gold coating facing the gas. As question of visible light tightness of the piston aroused (see Sec.3), transmission of the piston at λ=527 nm has been measured over a diameter of 100 microns, and found to vary between $1\ 10^{-5}$ to $2.5\ 10^{-5}$ from place to place. This variation in transmission may be the signature of local weakness or of tiny holes in the gold overcoat, more probable than 8% thickness variation that would also gives such variation in transmission. The other end of the tube is obturated by an XUV window, made from 0.1 to 0.2 microns of Al, SiC or $Si_3N_4$ and allowing backside XUV spectroscopic study in front of the radiative shock. Results obtained with the XUV spectrograph are discussed in other papers [13,16] and are not presented in this paper.

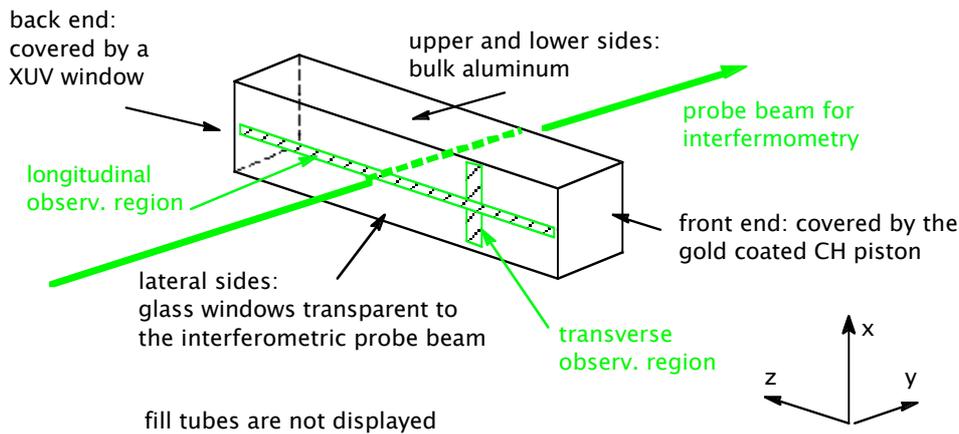

*Figure 2 - schematic layout of targets for interferometry. Selection of the observation regions (hatched areas) is done by rotation of the streak camera and its entrance slit. All 4 sides can be aluminium or gold coated.*

The PALS laser, focalized on the foil hereafter called piston, ablates the polystyrene. Following the ablation, the piston is accelerated by rocket effect and launch a strong, radiating shock in the Xenon gas [9]. The gold coating of the piston contributes to half of the inertia of the piston, but its first use is to block preheating of the Xenon gas by hard x-rays originated from the laser interaction region. However, experimental results suggest that the back layers of the gold coating may undergo a small preheating.

In front of the pure hydrodynamic density jump, a radiation driven ionization wave is launched: the radiative precursor. Electron density increase in the radiative precursor is due to increased ionization, following the increased temperature, but the ion density is unchanged.

Typical profiles computed with the one-dimensional hydrodynamic code MULTI [18,19] are displayed in Fig.3 .

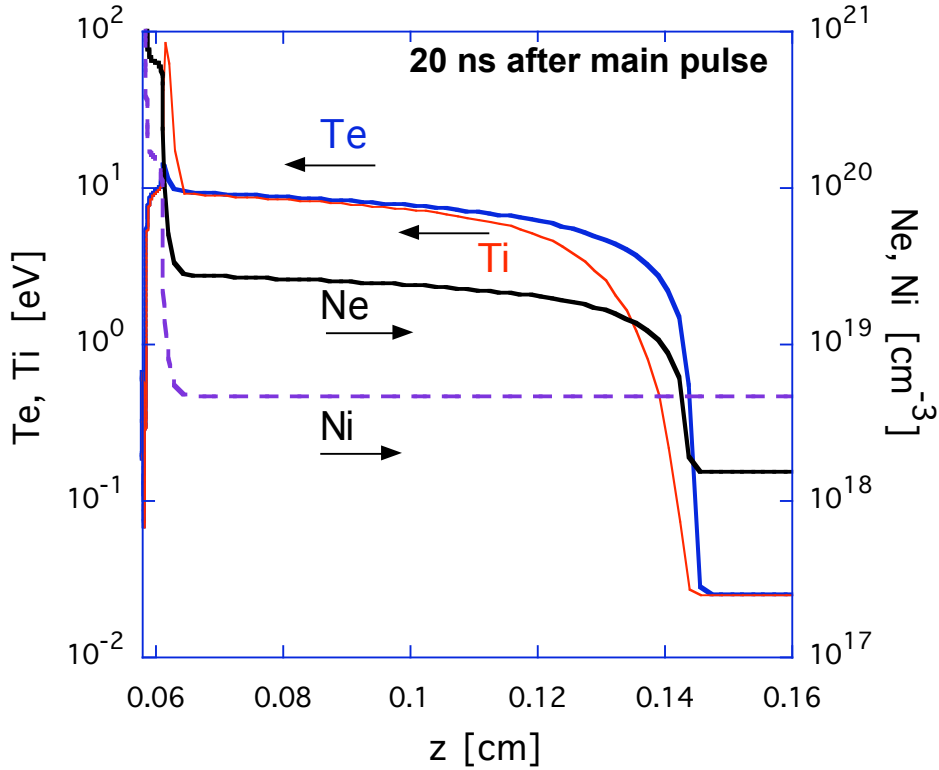

*Figure 3: Temperature and density profiles computed with the one-dimensional hydrodynamic code MULTI [18,19]*

For the time resolved interferometric diagnostics, we use an auxiliary laser working at $\lambda_P$=527 nm, with a pulse duration of ~ 300 ns. The probe beam is expanded to cover more that the tube length. The target is located in one arm of a Mach-Zehnder interferometer. The recombined two beams of the interferometer are imaged on the slit of a micro-channel plate enhanced streak camera with a CCD imaging camera (512 x 512 pixels). Interferometric filter at $\lambda$=527 nm and color glass filter located in front of the slit prevent any parasitic contribution from the plasma self-emission or from stray light from the main laser. Two configurations are used (Fig.2): (i) In the longitudinal configuration, we image the symmetry axis of the shock tube on the slit of the camera in order to record the shock propagation in the tube. (ii) In the transverse configuration, we image a section of the tube, perpendicular to the canal axis, to monitor the transverse shape of the shock wave by recording its time of arrival versus distance x from axis, at some fixed distance $z_0$ from the piston initial position. The observed regions in the two cases are displayed in Fig.2 by crosshatched areas.

For this interferometric diagnostic, we use targets with various parameters : in the first type of targets studied in this paper, labeled CF7 and CF20 in Table I, the shock tube is limited by gold coated aluminium on two sides of the square section and by glass optical windows on the two other sides, coated with a 15 nm lining of gold. The glass windows allow interferometric diagnostics with the PALS auxiliary green laser as a probe beam. The internal section of this 6 mm long square channel is 0.7 x 0.7 mm$^2$. The two other targets presented here, labeled CF61 and CF59 are similar to the previous ones, but with a wider internal section of 1.2 x 1.2 mm$^2$, with 2 bulk aluminium sides and 2 aluminized glass windows. They have been used for "transverse" interferometry. The shock propagates in Xenon with aluminum limiting walls for these last two targets CF61 and CF59, whereas it propagates between gold walls in the first targets CF7 and CF20. The nature of the channel coating affects the albedo and thus has an impact on the radiation energy lost from the shock and therefore on the velocity of the precursor front [11,20,21]. Actually, regarding radiation transport, the wall material acts as pure gold or as alumina ($Al_2O_3$). Specific experiments to access direct measure of the albedo have been performed since and are presented elsewhere.[22] We present measures of the electron density profiles in the next two sections.

| Shot Name | Target Name | Section shape | Transverse size | Length | Walls coating | Fill gas | laser energy |
|---|---|---|---|---|---|---|---|
| 1029_07 | CF7 | square | 0.7 mm | 6 mm | gold | Xe, 0.1 bar | 166 J |
| 1030_05 | CF20 | square | 0.7 mm | 6 mm | gold | 50-50 Xe+Air, 0.17 bar | 173 J |
| 1101_02 | CF61 | square | 1.2 mm | 6 mm | $Al_2O_3$ | Xe, 0.1 bar | 151 J |
| 1101_03 | CF59 | square | 1.2 mm | 6 mm | $Al_2O_3$ | Xe, 0.1 bar | 169 J |
| 1029_08 | CG32 | circle | 0.7 mm | 2.8 mm | $Al_2O_3$ | Xe, 0.1bar | 168 J |
| 1029_04 | CG35 | circle | 0.7 mm | 2.8 mm | $Al_2O_3$ | Xe, 0.1bar | 170 J |
| 1030_06 | CG50 | circle | 0.7 mm | 2.8 mm | gold | 50-50 Xe+Air, 0.17 bar | 171 J |

*Table 1 : summary of the main characteristics of the different shots and targets.*

## 3 – LONGITUDINAL INTERFEROMETRY

In the longitudinal configuration, time resolved interferometry traces the variation of the optical refraction index n(λ), at the wavelength λ of the probe laser, versus the position z on the cell axis along the direction of the shock propagation, and integrated over the path of the probe laser (y direction) along the cell width. In plasmas, the variation of the optical index, $n(\lambda)$, is usually attributed to free electrons, and is function of the electron density $N_e$,

$$n(\lambda) = [1-N_e/N_{c,\lambda}]^{1/2}$$

$$\sim 1 - 0.5\, N_e/N_{c,\lambda} \qquad (1)$$

where $N_{c,\lambda}$ is the critical density at the wavelength $\lambda$ of the probe laser. Thus, interferometry allows to measure the electron density $<N_e>$ as a function of the position z and time t, averaged over the transverse section d of the plasma

$$<N_e(z,t)> = \int_{canal\ width} N_e(y,z,t)\, dy/d \qquad (2)$$

where "z" is the distance from the initial position of the piston along the direction of propagation and "y" is the transverse position in the direction of the laser probe beam.

In the longitudinal configuration, we set the slit in front of the camera to image only the very center of the tube. Slit width was generally ~100 microns.

**3-1 qualitative analysis**

We first present in this section results obtained with the interferometric diagnostic for the target CF7, shot #1029_07, filled with Xenon at 0.1 bar. The "post mortem" analysis of the target indicates a correct centering of the main laser impact. The x-ray pinhole image shows a smooth distribution of the laser intensity on the target. Raw image and processed interferogram are displayed in Fig.4 . The probe intensity profile (derived from the record at late times) is substracted to the raw image without (Fig.4b) or with additional FFT filtering removing high spatial frequency (as used in Fig.9). The vertical axis is the time running for 100 ns from the bottom of the image. In this particular shot the laser pulse arrive just at the beginning of the sweep. The horizontal axis is the distance z from the initial position of the piston, located on the left of the figure. The 6 mm length of the canal is imaged on the whole slit and correspond to the full width of the image. The plasma plume ablated from the piston front side is not imaged in the observed spatial range. The horizontal bean-like dark areas are burnt pixels of the CCD. Low frequency filtering can remove them, but would smear out broadened fringes (around z=1.5mm and t=10ns) as well as pattern shifted by one half-period (around z=4.5mm and t=25ns).

Initial fringe shift, in zone "a" (lower part of Figs.4,5) at early times is attributed to electromagnetic perturbation to the camera electronic tube due to the plasma breakout. Therefore reference position of the fringes will be taken after the perturbation. The interferogram presents two distinct zones of moving fringes. In zone "d" (t=40 to 60 ns, z=3 to 4 mm), the inclined fringes pattern is the signature of an electron density gradient, and is attributed to the radiative precursor [9]. The precursor foot (line "A"), where the fringes start to shift, is recorded at an average velocity around ~85 km/s, decreasing from an initial velocity of 120±20 km/s to a mere 60 km/s 4 ns later.

On the lower right (z>4.5mm) another front is seen, with sub-fronts detected (lines Gx), moving away from the rear of the miniature shock tube. This is attributed to another radiative shock (with a velocity of ~30km/s decreasing to ~10 km/s) launched by some faint preheat of the rear side window. The different patterns of the shifted fringes correspond to the sign of the density gradient as discussed in Koenig et al. [23]. Various interpretations can be proposed for the heating of the back window within a few ns from the laser pulse: fast electrons, suprathermal x-rays, residual laser light going through or around the piston. Preheating by fast electron occurs usually at higher laser intensities than the nominal value of our experiments ($10^{14}$ W/cm$^2$). Suprathermal x-rays are stopped by the gold layer and would not transport enough energy to ablate the back window. Suprathermal electrons or x-rays would probably equally preheat inner sides of lateral windows and back side window, whereas recorded interferograms evidence no perturbation before the direct and reverse radiative waves have traveled from their birth points. Furthermore, heating of lateral windows would squeeze the free channel, whereas heating of the back window launch an upstream wave. Because of the chromaticity of the focusing lens, the unconverted infrared (1ω) beam has a transverse size of 1 cm$^2$ at the target location, compared to the 0.5 mm$^2$ section of the tube. With less than 3J of unconverted light, it results that only a few mJ of infrared might go inside the tube and cannot launch an ionization wave. Heating from a fraction of Joule of residual converted (green) laser light going through tiny holes of the gold coating, or a few J in the wings of the PZP image going around the piston, is more likely. We have checked with an improved version [19] of the hydro code MULTI [18] that 0.1-0.2% of the incident energy can heat the end of the tube and launch a radiative wave. However, it was not possible to account for the exact complex geometry of the tube end, probably responsible of the sub-fronts launch.

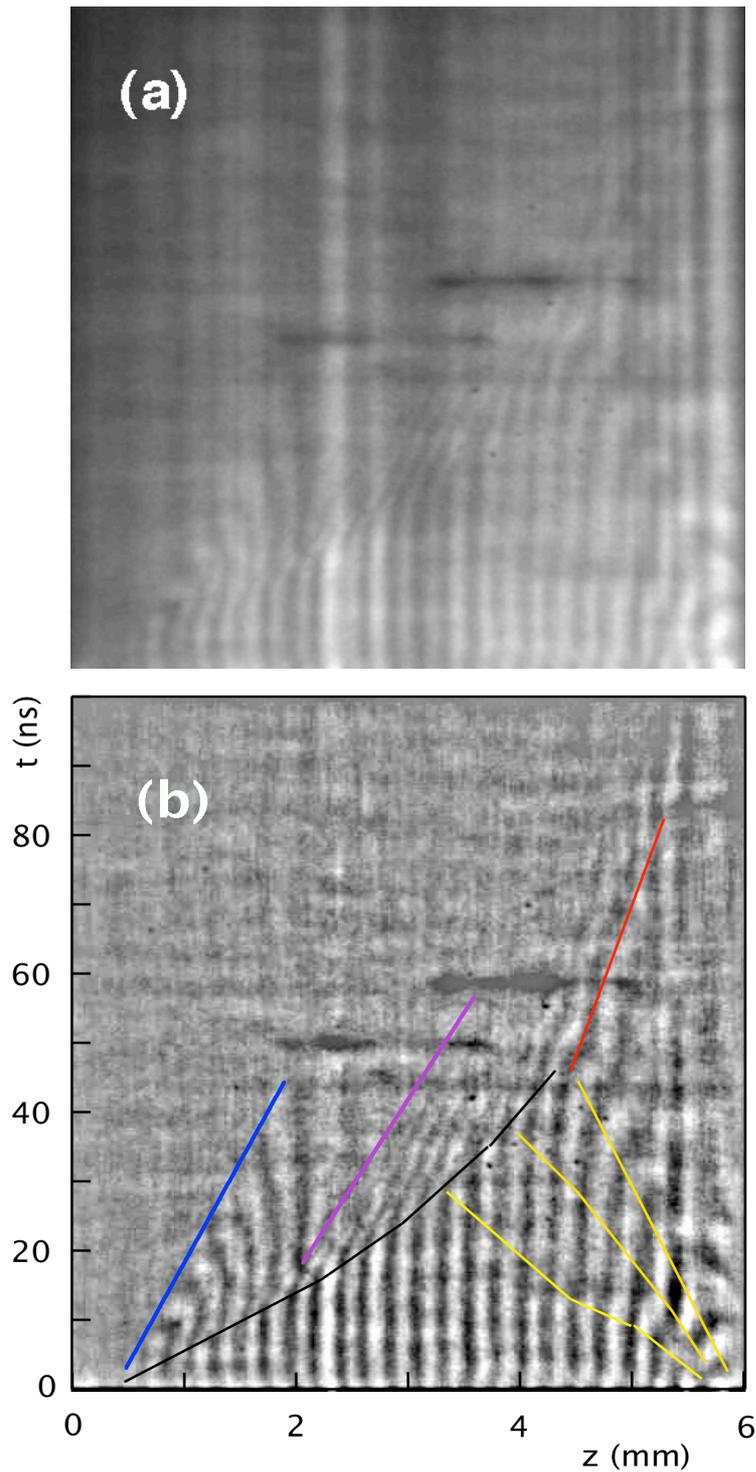

*Figure 4: shot #1029_07 interferogram in longitudinal configuration for a target filled with Xenon at 0.1 bar. Raw image (a) and processed image (b) with back ground substraction. Additional FFT filtering is used in Fig.9. Time runs on vertical axis for 100 ns, distance z from the initial position of the piston runs on horizontal axis for 6 mm. Main shock goes from left to right. Superimposed lines identify the different time and space domains discussed in the text.*

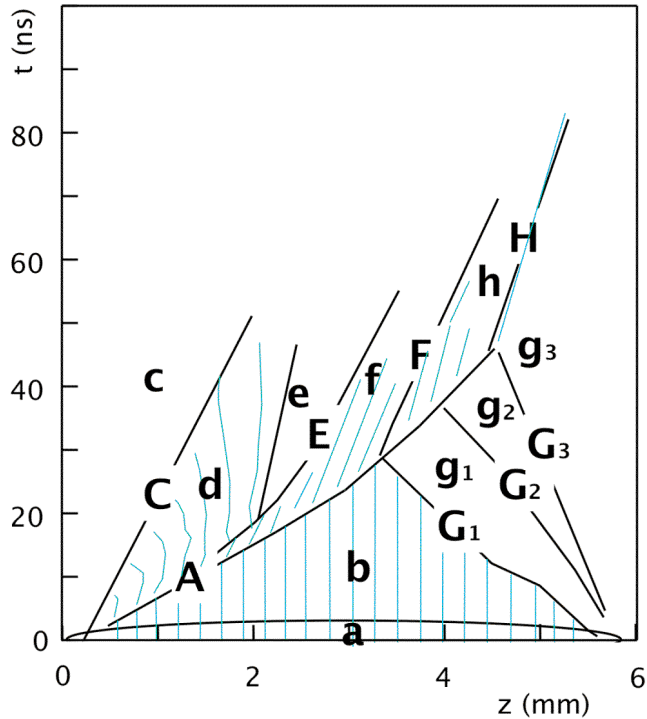

*Figure 5: full identification of zones and limits observed in Fig.4: zone a: electro-magnetic perturbation to the streak camera at early times; zone b: unperturbed space & time region; zones d,e,f,h: radiative precursor, precursor front is located on line A; zones g: reverse radiative precursor; zone e; splitting of radiative precursor; zone f: fully developed ionization wave; zone h: precursor slows down as it propagates in the plasma preheated by the upward ionization wave; zone c: shadow of the piston, we believe that the rear surface of the piston is located on line C, moving at a velocity around 30 km/s. All these features have been identified on other shots, with different laser and target conditions.*

In the leftmost part of zone "d", we observed that the fringes, initially shifting to the right, start to shift back, following the decrease with time of the electron density: fringes inclined to the right (resp. to the left) means increasing density (resp. decreasing) with time. This increasing-decreasing density feature is associated with preheating of the gold layer on the back side of the piston, inducing a small heat wave in the gas, as figured out from MULTI simulations. This observation comforts our confidence that the position of the piston is located at the end of the observable fringes (line C) after the obscure region "c" where the electron density in front of the piston (i.e. facing the laser) is large enough to absorb consequently the probe beam and to shift fringes more than observable. Synthetic interferogram (see discussion in the last paragraph of this section) confirms this analysis.

The more or less flat area "e" is understood as a split precursor, when an ionization wave is progressively detached from the main precursor (Fig.6). This feature has not been seen with MULTI simulations using gray radiative transfer, as it is probably the results of spectral dependence of the radiative transfer. Multi-group, multi-dimensional numerical simulations will be carried out, using new methods in the HERACLES [25] code (implementation in progress) to address this question. Note that ahead of a diffusion radiation wave (a Marshak wave), there is generally a small foot, an exponentially decaying free-streaming region created by the high energy tail (relatively to the local temperature) of the photon distribution. This foot is the seed of the diffusive wave, well distinguished by the concavity of the temperature profile (downward for the diffusive wave, and upward for the seeding foot), whether the diffusive wave remains single or is progressively split in two because of steep variation with energy of the photon mean free path. An exponential decaying foot does not produce this plateau feature seen in the interferogram. At this time, we have no definite interpretation of this split precursor, but only a clear unambiguous observation of the phenomenon.

All these features have been identified on other shots, with different target conditions. In shot #1030_05 (Fig.7), the fill gas was a 50-50 mixture of Xenon and air , pressure was 0.17 bar. It displays a split precursor signature (V-shaped flat fringe between 2 zones of parallel fringes with different slopes), and a backwards precursor coming from the bottom of the tube. Slow down of the main precursor after shock crossing is equally clearly seen.

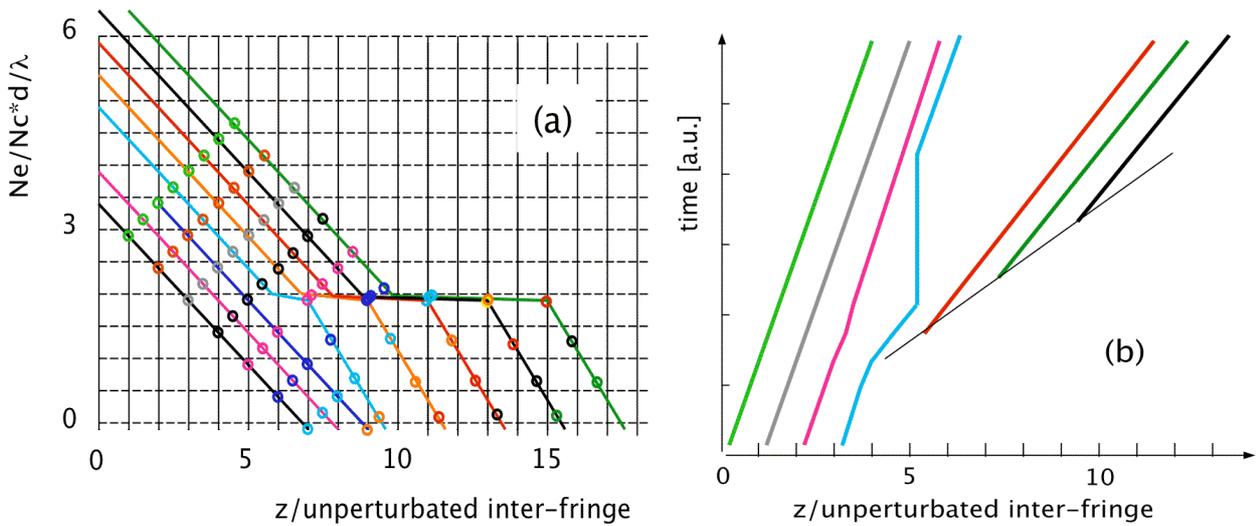

*Fig.6 fringes positions (b) derived from schematic density profiles (a) at successive times with an ionization wave detaching from the initial precursor.*

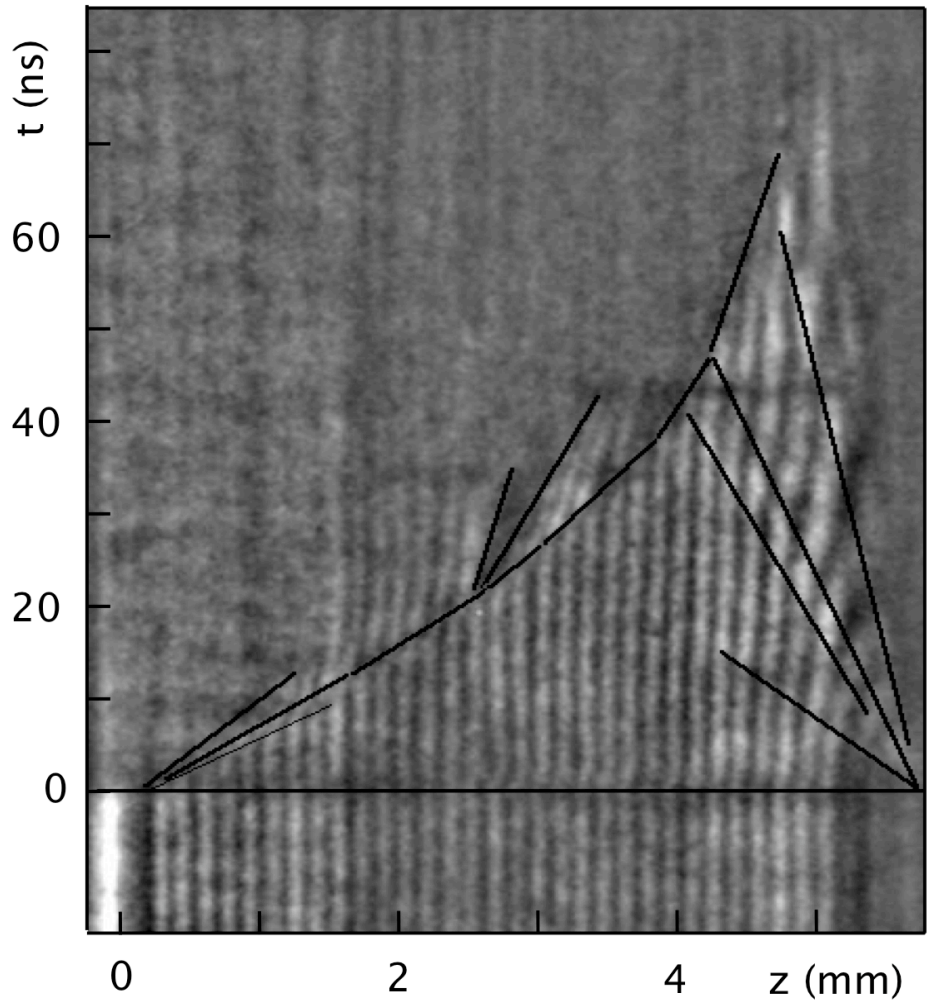

*Fig.7 processed image of shot #1030_05 (50-50 Xenon/air mixture at 0.17 bar) also displays a split precursor signature (V-shaped flat fringe between 2 zones of parallel fringes with different slopes), and a backwards precursor coming from the bottom of the tube. Slow down of the main precursor after shock crossing is equally clearly seen.*

In Fig. 8, we plot a synthetic interferogram derived from the time history of the electron density computed with our extended version [19] of the MULTI code. For this simulation, we use a laser intensity of $10^{14}$ W/cm$^2$ on piston and $10^{11}$ W/cm$^2$ impinging the rear window. It reveals qualitatively the same zones than the observed interferogram. Although still a one dimension hydrodynamic code, our version of MULTI includes simple modeling of 2D effects with an approximation of the lateral radiative losses [24]. The computed final slope of the precursor is sensitive to the wall albedo used in the simulation. Here we used 60% with an equivalent diameter of 700 microns. When including a fraction of Joule deposited in the gold layer of the piston, we even observe (Fig.8b) in the synthetic interferogram the commas seen in zone "d", revealing increase of electron density followed in time by some density decrease, in Xenon close to the piston. Exact structure of this energy deposition is not known, so precise shape of these comma features has not been reproduced. The split precursor region is not reproduced in this simulation. It probably requires multi-group radiation transfer to show up.

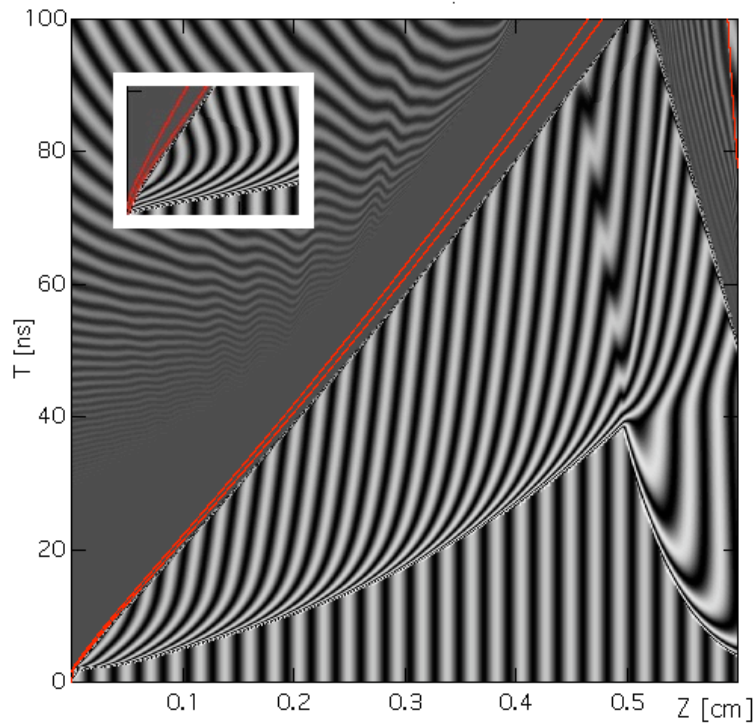

*Figure 8: Synthetic interferogram (a) with simple modeling of 2D effects of the lateral losses [24], computed for 135 J laser energy on piston, and 0.1 J on back window (we found other combinations giving similar results). The exact shape will depend on 2D effects and on ionization value around 10 eV. As seen in inset, including a 0.3 J preheat in the gold back layer of the piston creates comma-shaped fringes as observed in zone "d" of experimental record.*

## 3-2 quantitative analysis

Using the processed image (Fig.9) where we plot the position of the fringe maxima, we are able to follow the fringes displacement versus time. We can then derive the contour lines of the electron density as the phase shift follows the average electron density over the d=0.7 mm width of the tube :

$$\varphi(z,t) = 2\pi \frac{d}{\lambda} \int_0^d \left(1 - \sqrt{1 - \frac{N_e(y,z,t)}{N_{c,\lambda}}}\right) dy$$

$$\approx \pi \frac{d}{\lambda} \int_0^d N_e(y,z,t)\, dy \,/\, N_{c,\lambda} = \pi \frac{d}{\lambda} <N_e(z,t)> \,/\, N_{c,\lambda} \quad , \qquad (3)$$

where $\lambda$ is the wavelength of the probe laser (here $\lambda$ = 527 nm) and the average (for a given z) electron density $<N_e(z,t)>$ is given by Eq. (2).

The precursor front, taken at the position where the fringes are shifted by 1/10 period corresponding to $<N_e>$ = 7 $10^{17}$, shows a deceleration with time. Initial velocity is larger than 130 km/s, slowing down to ~90km/s after the full buildup of the precursor (between 2 and 3.5 mm from initial position of the piston) and then around 60 km/s. The reconstructed contour lines of the electronic density $<N_e(z,t)>$ (averaged over the width of the canal) are reported in Fig. 9 for values between 0.7 $10^{18}$ and $10^{19}$ cm$^{-3}$. Fringes have almost the same slope between 20 and 50-60 ns, a signature of a stationary gradient length. At this time, the precursor front velocity is measured around ~90 km/s. Previous studies [11,20,21] have shown that the lateral radiative losses have an important role in the shock structure and dynamics. In particular, the precursor velocity is much slower when there are lateral losses, and slows down until it has reached a stationary limit with then a velocity equal to the velocity of the hydrodynamic shock (i.e. density jump). This limit has not been reached in our experiment before the main precursor crosses the reverse precursor coming from the end of the tube (lines Gx in Fig. 5). On the contrary, we observe in the recorded interferogram that the gradient region (zone "f" in Fig.5) increases with time. Another point is that the main precursor is observed to slow down after crossing the reverse wave (at z=3.5mm). Slowdown of the precursor front after crossing with the reverse radiative precursor has been observed in other shots. As the mean opacity changed with temperature, the change of the optical properties of Xenon heated by the reverse wave is expected to modify the precursor front velocity.

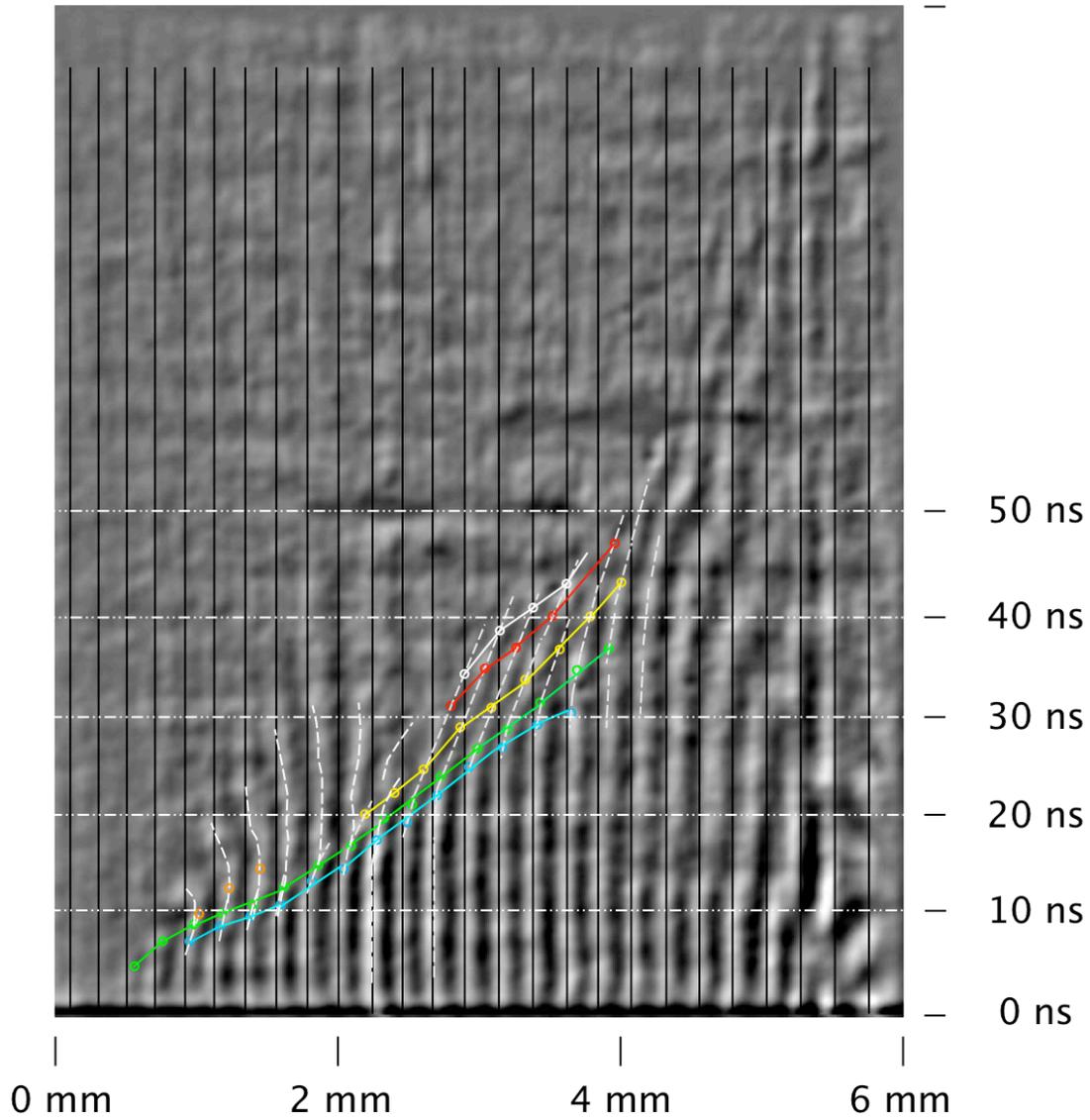

Figure 9 : Contour lines of fringe shift (which are also contour lines of electron density) superimposed to a processed interferogram, corresponding to $N_e/10^{18}$ cm$^{-3}$ =0.7, 2, 5 and 10 (in blue, green, yellow and red).Vertical and horizontal scales are 100 ns and 6 mm. Superimposed black vertical lines stands for the unshifted position of the fringes after the initial perturbation. White dashes enhance the fringe centers.

We numerically studied the shock 2D properties for different values of the lateral radiative losses. The boundary condition was of a constant piston velocity of 63 km/s, corresponding to the result of a MULTI 1D simulation of the whole target. Although the piston velocity is measured to be 30 km/s, the preheating of the gold back layer is equivalent to an increased piston velocity. MULTI simulations, including lateral radiative losses, yield similar precursor front with a 30 km/s velocity preheated piston or with a 63 km/s velocity cold piston. 2D results presented below suggest also that it is a good estimation. The 2D simulations are performed with the HERACLES hydrodynamic code [25], which includes M1 radiation transport, Super Transition Array opacity [26] and equation of state from OPA-CS hydrogenic model [27]. We report in Fig. 10 the computed and measured average transverse electron density in the precursor at two different times (20 and 40 ns) and for two values of the lateral radiative losses (20% and 40% of incident x-ray fluxes) at the gold walls of the tube. From this comparison we can infer a value of lateral losses around 35% (albedo=0.65). As expected from the material albedo (albedo increases with the atomic number), this value is smaller than the value of 60% derived from earlier studies where the shock was propagating in a glass and aluminium tube [11], instead of a gold-coated tube in the present study. The computed gradient length $L_\nabla$ of the precursor front presents a small time variation, in agreement with the observation of almost parallel fringes in the recorded interferogram between 20 ns and 40 ns (Fig.10), after the precursor built up and before the upstream precursor crossing.

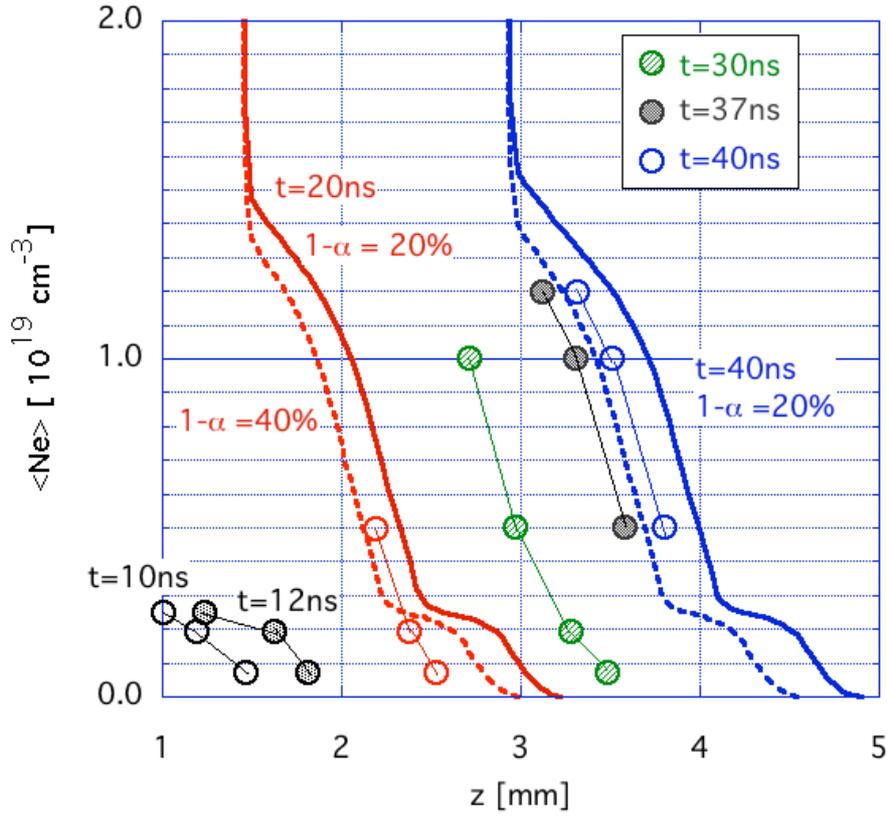

*Figure 10 : transversally averaged electron density <Ne (z,t)> at two times (20 ns in red and 40 ns in blue) : experimental values (circles) and electron densities computed with HERACLES for two values of the lateral radiative losses (dashes: 40 % , solid line: 20 % ). Note that the simulation shows a quasi-constant gradient length.*

We apply the same filtering procedure to the interferogram from shot #1030_05 (displayed in Fig.7) and derive the following plot of density profile (Fig.11) for time between 10 ns and 40 ns after the laser pulse. In this shot, the shock tube was filled with a 0.17 bar, 50-50 Air-Xenon mixture. *This shot exemplifies the split precursor* with its plateau signature. The precursor moves with a velocity of ~ 90 km/s and the density peak velocity is around 60 km/s. Precision is estimated around $5\ 10^{17} cm^{-3}$ and 0.01 mm. Note that the density profiles in the density peaks drawn with dashes in Fig.11, are indicative and *cannot* be measured from the interferogram with the interferometer setting we used.

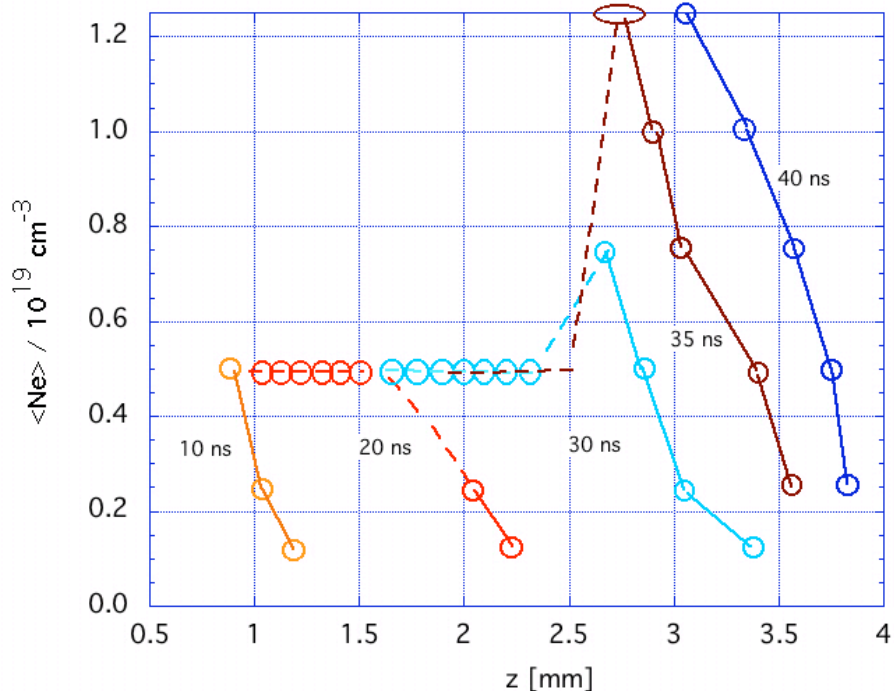

*Figure 11: electron density profiles at different times after the laser pulse derived from the interferogram displayed in Fig.7. Precision is estimated around $5\ 10^{17} cm^{-3}$ and 0.01 mm. The dashed lines are drawn for guidance but lay in the V-shaped region where density cannot be measured.*

## 4 – TRANSVERSE INTERFEROMETRY

Longitudinal interferometry gives information about the time variations of the averaged transverse electron density $<N_e(z,t)>$, versus the direction z of shock propagation. It gives the longitudinal structure of the radiative precursor, and allows measure of the front velocity, but it does not give *direct* information about the 2D shock structure. To access it, we performed a transverse interferometric study with the same kind of target, turning the camera by 90°. Now, the slit of the streak camera images a transverse section of the shock tube at a distance $z_0$ from the initial position of the piston. Transverse interferometry traces time variations of the electron density $<N_e(x,t)>$, averaged over the other transverse direction y, at a fixed distance $z_0$.

For this study we used targets CF61 and CF59 with no gold coating but aluminium coating. The gas in the tube is then in contact with aluminium walls on two opposite sides and with aluminized glass windows for the two other sides. Actually, aluminium rapidly oxidizes into alumina ($Al_2O_3$) and oxygen lines have been observed with the backside XUV spectrograph [13,16,17]. The section of the cell, 1.2 x 1.2 $mm^2$, is larger than the nominal focal spot (~ 0.7 mm), therefore curvature of the shock front from pure geometrical reasons is expected in the beginning (~1mm) of the propagation. The tube length is again 6 mm. These targets are filled with Xenon at 0.1 bar.

The results are presented in Fig. 12 for two different values of the distance from the piston, $z_0$=1.5±0.5 mm and $z_0$=3.0±0.5 mm. In the latter case, the canal width was obscured by some misaligned mask, and only 1mm is visible, leaving a 0.2 mm obscure zone on the left of the interferogram (Fig.12b), not noticed in preliminary results [13]. The horizontal axis corresponds to the transverse direction (1.2 mm) and the vertical axis to the time (100 ns).

Post mortem analysis of the targets shows that, in the first case (a), the laser pointing was very close from the lateral wall (approx. one half of the focal spot actually in front of the canal) and fully inside the canal section for the second (b), although shifter from center by 0.15 mm, as deduced from the interferogram. Laser energy in these two shots is ~150 J and ~170J. Probably because of a hiccup of the laser, the main pulse delay after the camera synchronizing pulse is 20 ns later on shot #1031_02 (Fig. 12b) than in shot #1101_03 (Fig.12a). The small initial "collective" shift mentioned earlier can be used as a time fiducial of the main pulse. Note that a confirmed double pulse in another shot results in the same time delay. We have checked with hydrodynamic simulation that double pulse does not change the structure and velocity of the radiative shock launched in the Xenon, as a consequence of the launch by rocket effect. Only longer pulse and ramp shape would significantly change it. The positions of laser time arrivals are reported with black arrows, and the location of the focal spot is displayed above the interferograms. The lateral extension of the flat part (i.e. with same arrival time) of the shock section is around 200 microns at t=30 ns (Fig.12b), and approximatively 400 microns at t=17ns (Fig.12a) if we assume that the focal spot is centered on the edge of the canal.

As the width of the canal exceeds the shock section, the radiative losses are increased. Moreover, lower piston velocity at the edges of the focal spot (because of lower laser intensity) –and beyond– is expected. As a consequence, a spherical shock behavior is expected until the first 1-1.5 mm distance of propagation before recovering a somehow planar structure by filling the full width of the canal. On the contrary, lateral radiative losses will cool down the shock temperature peak close to the walls and later on even create some curvature of the precursor and the shock itself [21]. 2D simulation is then required for full understanding of the front shapes.

Recorded interferograms (Fig.12) show indeed lower shock front velocities far from the center of the focal spot, or more precisely, arrival of higher density and/or steep gradient appears later far from the center. The time delay $\Delta t$ between the precursor arrival at x=500 µm and x=0 from the focal spot center is measured to be $\Delta t \sim 10$ ns (resp. $\Delta t \sim 5$ ns) after a propagation distance of $z_0 = 3$ mm (resp. $z_0 = 1.5$ mm). On the other hand, the precursor foot, identified by the beginning of the fringe shift (shown as a straight white dashed line in Fig.12), appears already planar (arriving at the same time on the whole width of the canal) for the two distances from piston we observed. Such a feature of a flat precursor foot ahead of a curved shock has been observed numerically with the 2D hydro-rad simulation by Leygnac et al, at least at early time, see Fig.9 of ref. [20]. They use a realistic laser intensity distribution in the focal spot, close to the one we have in our experiment, and a comparable transverse size of the channel.

The times of disappearance of the fringes at the center of the focal spot, and on the left of the canal in shot #1101_03, and at approx. 0.15 mm from the edge in shot #1101_02, are respectively ~13.5 ns and ~31 ns. However, distances from the piston are known with a poor accuracy, not better than 0.5 mm. This corresponds to an average velocity around 110±20 km/s and 95±20 km/s from 0 to resp. 15 ns and 35 ns. These value are compatible with the precursor velocity derived from longitudinal interferometry at these times of 15 ns and 35 ns, although we expected smaller values due to increased lateral losses.

We note also that in both records, the precursor foot, when the fringes start to shift, arrives at the same time in the whole width of the canal. Translated in terms of spatial shape, the precursor foot is already planar, though the precursor breakout, when the fringes disappear, shows largest delay in the wings of the focal spot, which translates into a spatially curved front.

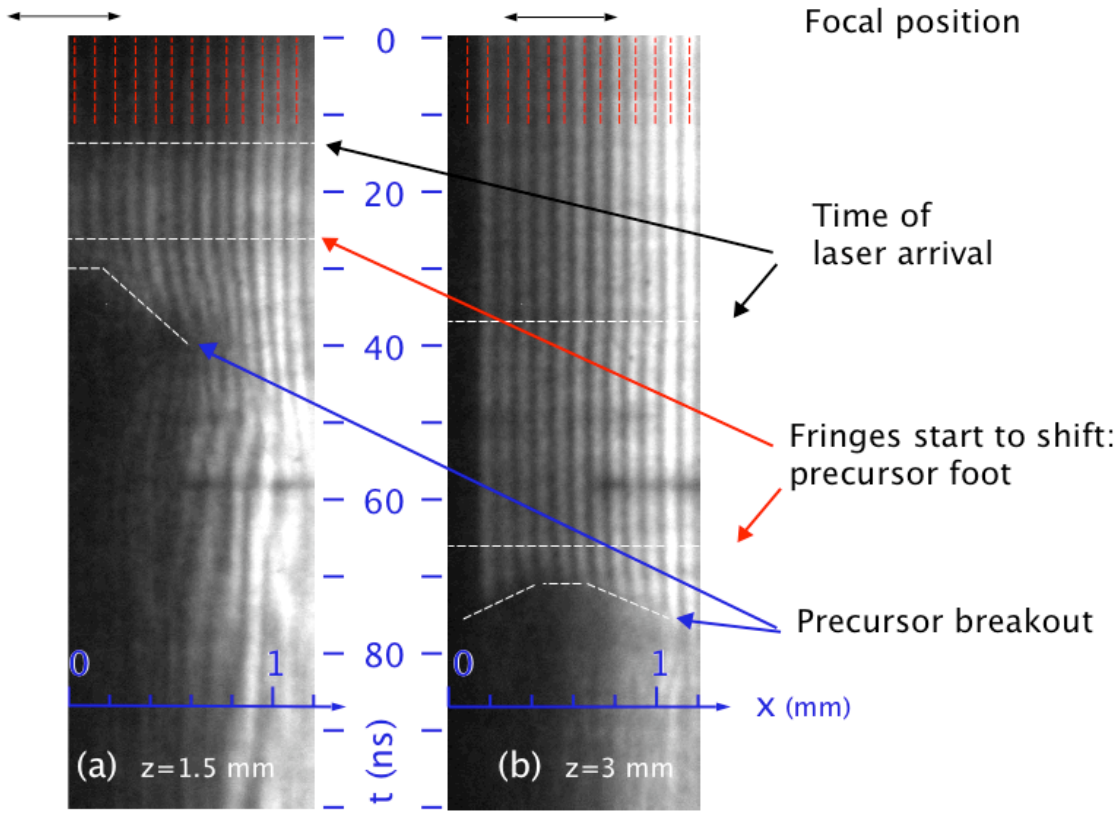

*Figure 12 : "transverse interferograms" of 1.2 mm wide canals at two different distances $z_0$ from the initial piston positions, approximatively 1.5 mm (a, shot #1101_03) and 3.0 mm (b, shot #1101_02). The field of view is reduced to ~ 1 mm width in (b). Time runs vertically for 100 ns (top to bottom) and transverse position runs horizontally for 1.2 mm. Laser time arrival is pointed with black arrows, at 13.5 ns (a) and 36 ns (b), while precursor foot is pointed with red arrows. The horizontal arrows on top of the interferograms correspond to the focal spot positions.*

Using the same procedure than for "longitudinal" interferograms, we are able to derive the time evolution of the electronic density at this distance $z_0$=3 mm from the piston, and at lateral distance x from the focal spot center between 0 and 0.6 mm. These time profiles are displayed in Fig.13. The rise time are almost equal on the whole width, but with a larger delay far from center. Except for the very beginning which appears to arrive at the same time, as already mentioned, the time history for each distances from center are very close in shape, but with a variable delay increasing from center to edges. Translated in terms of propagation distances, this means that the precursor length is rather constant on the whole width of the tube, but that the precursor presents some curvature, again except for the very beginning of the precursor which is almost flat.

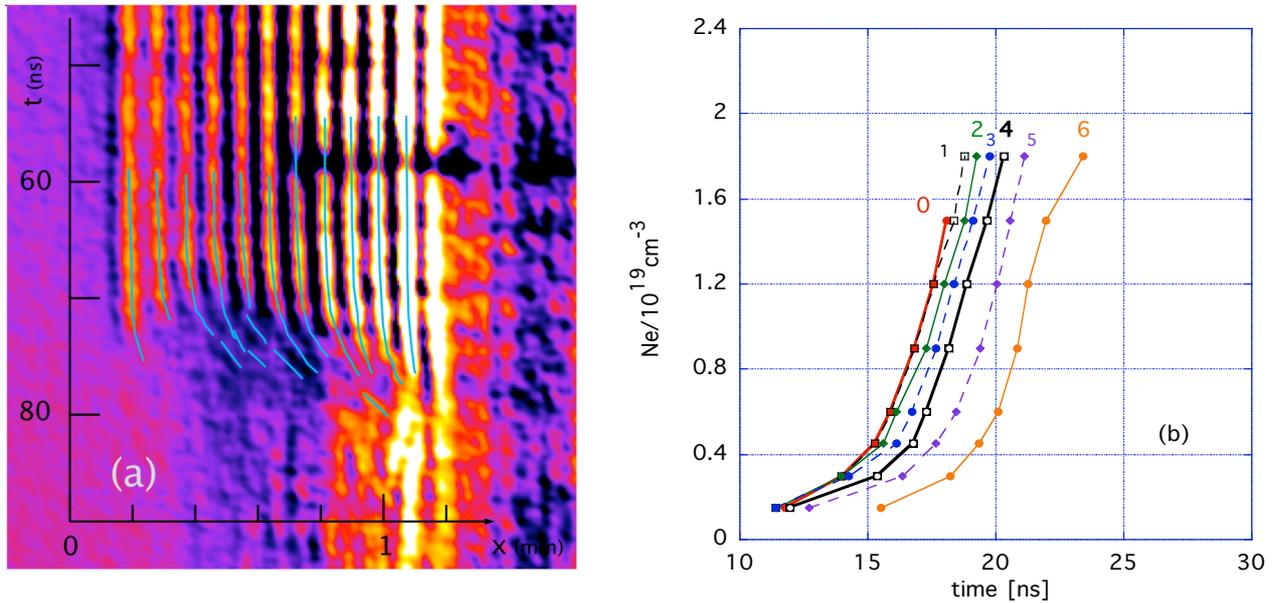

*Figure 13 : transverse interferometry at distance z=3mm (shot#1101_02): (a) processed zoomed interferogram and (b) time evolution of electronic density at different distances from center. Numbers on top of curves are distance from the focal spot center in units of 0.1 mm. Misalignment of some mask limit the visible part of the canal to 1 mm.*

Pure hydrodynamics tell that the shape of the hydrodynamic shock (i.e. the density jump) should be flat after propagation over a few diameter of the shock tube. However radiative losses through the walls have been demonstrated to create a curved precursor [11, 21], in competition with the hydrodynamic flattening. 2D simulations are needed to discriminate between transverse variation of the piston velocity, spherical propagation, effect of lateral radiative losses and address competition between these effects.

## 6 – CONCLUSIONS

In this paper, we present comparison with 2D simulation of the recorded traces of a radiative precursor with an average front velocity around 85 km/s. Though there are puzzling questions remaining, measured electron density profiles at different times are in agreement with simulation. The results are consistent with radiative losses at the gold-coated walls of 35% (albedo of 0.65), expectedly less than the 60% for Al (albedo of 0.4) derived in the previous experiment [11]. Two other interesting features have been observed, a slowing down of the radiative precursor in pre-shocked gas (heated by an other, upstream, radiative wave), and a detached ionization wave or split precursor. In the shots we analyzed, the stationary regime has not been reached before the precursor crosses the reverse radiative wave coming from the other end, though we observe an almost stationary gradient length of the main ionization wave. On the contrary, we observed that the extension of the gradient region increases with time. Transverse analysis of the radiative precursor has also been done, and gives a precursor velocity in agreement with the velocity derived from the longitudinal analysis. Time history of electron density at various distances from the tube axis has been measured at 35-40 ns after the laser pulse. Further 2D simulations are required for full analysis.

## 7 – ACKNOWLEDGMENTS

The experimental campaign "PALS project#1077&1386" reported in this paper was carried out in 2007, involving Ouali Acef, Patrice Barroso, Michel Busquet, Norbert Champion, Chantal Stehlé (Observatoire de Paris), Guillaume Loisel, Frédéric Thais (CEA Saclay), Krzysztof Jakubzak, Michaela Kozlova, Tomas Mocek, Jiri Polan, Bedrich Rus (Department of X-Ray Lasers, Institute of Physics / PALS Center), under the P.I.ship of Chantal Stehlé and Michel Busquet in the framework of the LASERLAB facilities network [28]. The authors give a *special thank* to P.Barroso and O.Acef whose contributions were instrumental in the fabrication of the targets and in the design and assembly of the Mach-Zehnder interferometer. We thank O. Madouri (LPN) and V.Petitbon (IPN) for providing us with the SiC windows and the gold coated polystyrene thin films. We acknowledge financial supports from the Access to Research Infrastructures activity in the Sixth Framework Program of the EU (contract RII3-CT-2003-506350 LaserLab Europe), from the RTN JETSET (contract MRTN-CT-2004 005592), from the French National Program of Stellar Physics (PNPS) CNRS-PICS 4343. One of the authors (M.G.) acknowledges the financial support provided by the Spanish Ministry of Science and Innovation through the Juan de la Cierva grant. This research was partially supported by the Czech Ministry of Education, Youth and Sports (project LC528). We also acknowledge financial support from the OPTON Laser International Corporation.

# 7- REFERENCES

[1] Y.B.Zeldovich, Y.P. Raiser, (1966) Physics of Shock Waves and High Temperature Hydrodynamic Phenomena, New York: Academic Press; N. Kiselev Yu, I.V.Nemchinov, V.V.Shuvalov, Comput. Math. Phys., **31** (1991), 87-101

[2] L.Ensman and A.Burrows, *et al*, Astrophys. J. **393** (1992), 742

[3] P.Ghavamian, *et al*, Astrophys. J. **535**, 623 (1999)

[4] J. Bouvier, *et al*, "Magnetospheric Accretion In Classical T Tauri Stars, In Protostars And Planets" V, B. Reipurth, D. Jewitt, And K. Keil (Eds.), University Of Arizona Press 2007, Tucson, P. 479-494; J.F. Donati, *et al*, Mon. Not. R. Astron. Soc., **386**, 1324 (2008).

[5] B.Reipurth and J.Bally, Annu. Rev. Astron. Astrophys. **39**, 403 (2001) ; C.Feinstein, *et al*, Astrophys. J. **526**, 623 (1999) ; A.Raga, *et al*, Rev. Mex. A&A. **35**, 1304 (1986)

[6] S.Bouquet, R.Teyssier, J.P.Chieze, Astrophys. J. Suppl. Ser. **127** (2000), 245; C. Michaut, C. Stehlé, S. Leygnac, T. Lanz, L. Boireau, Eur. Phys. J. D, **28** (2004), 381-392

[7] J.Grun, J.Stamper,C. Manka, J.Resnick, R.Burris, J.Crawford, B.H.Ripin, Phys. Rev. Lett. **66** (1991), 2738

[8] C.A.Back, *et al*, Phys. Rev. Lett. **84** (2000), 274; D.Hoarty, *et al*, Phys. Rev. Lett. **82**, 3070 (1999); P.A.Keiter, R.P.Drake, T.S.Perry, H.F.Robey, B.A.Remington, C.A.Iglesias, R.J.Wallace, Phys. Rev. Lett. **89**, 165003 (2002)

[9] X.Fleury, *et al*, Laser & Particle Beams, **20** (2002), 263; S.Bouquet, *et al*, Phys. Rev. Let. **92** (2004), 5001

[10] A.B. Reighard, R. P. Drake, *et al*, Phys. Plasmas **13** (2006), 082901; A.B. Reighard, R.P. Drake, J.E. Mucino, J.P. Knauer, M. Busquet, Phys. Plasmas **14** (2007), 056504.

[11] M.Busquet, E.Audit, *et al*, High Energy Density Physics **3** (2007), 8; M.González, *et al*, Laser Part. Beams, **24** (2006), 535.

[12] J.C.Bozier, G.Thiell, J.P.Lebreton, S.Azra, M.Decroisette, D.Schirmann, Phys. Rev. Lett. **57** (1986), 1304

[13] C. Stehle, M. Gonzalez, M. Kozlov, B. Rus, T. Mocek, O. Acef, J.-P. Colombier, T. Lanz, N. Champion, K. Jakubzak, J. Polan, P. Barroso, D. Baudin, E. Audit, J. Dostal, and M. Stupka, "Experimental study of radiative shocks at PALS facility," Laser Part. Beams (in press).

[14] K. Jungwirth, Laser Part. Beams, **23** (2005), 177.

[15] I.N.Ross, D.A. Pepler, C.N. Danson, Optics Comm. **116** (1995), 55.

[16] M.Busquet, Patrice Barroso, Thierry Melse, Daniel Bauduin, Rev. Sci. Instr. **81**, 023502

(2010)

[18] R. Ramis, R. Schmalz, J. Meyer-Ter-Vehn, Comp. Phys. Comm. **49** (1988), 475.

[19] M.Busquet, An Extended Version of the MULTI Code, unpublished (2008).

[20] S.Leygnac, L.Boireau, C.Michaut, T.Lanz, C.Stehlé, C.Clique, S.Bouquet, Phys. Plasmas, **13** (2006), 113301; ibid, arXiv:astro-ph/0610876v2.

[21] M. González, E. Audit, C. Stehlé, A&A, **497** (2009), 27.

[22] M.Busquet, F.Thais, G.Ghita, D.Raffestin, Bull. Am. Phys. Soc. (2009); ibid., in prep.

[23] M. Koenig, *et al*, Phys. Plasmas, **13** (2006), 056504.

[24] M.Busquet, E. Audit, M. González, Bull. Am. Phys. Soc. **53**(14) (2008), 301.

[25] M.González, E. Audit, P. Huynh, A&A **464** (2007), 429.

[26] A.Bar-Shalom, D. Shvarts, J. Oreg, W.H. Goldstein, A. Zigler, Phys.Rev. A **40** (1989), 3183.

[27] C. Michaut, C. Stehlé, S. Leygnac, T. Lanz, L. Boireau, Eur. Phys. J. D **28** (2004), 381.

[28] http://www.pals.cas.cz/pals/projekty/proj1077.htm
http://www.pals.cas.cz/pals/pae104fr.htm
http://www.pals.cas.cz/pals/pae104ma.htm